\documentclass[final,3p,times]{elsarticle}

\usepackage{amssymb}

\begin{document}

\begin{frontmatter}



\title{Equation of state for all regimes of a fluid: from gas to liquid}


\author[add1,add2]{M. V. Ushcats}
\ead{mykhailo.ushcats@nuos.edu.ua}

\author[add1]{S. Yu. Ushcats}
\author[add2]{L. A. Bulavin}
\author[add2]{V. M. Sysoev}

\address[add1]{Admiral Makarov National University of Shipbuilding, 
 9, Prosp. Heroes of Ukraine, Mykolayiv 54025, Ukraine}
\address[add2]{Taras Shevchenko National University of Kyiv, 
 2, Prosp. Academician Glushkov, Kyiv 03680, Ukrain}

\begin{abstract}
The study of Mayer's cluster expansion (CE) for the partition function demonstrates a possible way to resolve the problem of the CE non-physical behavior at condensed states of fluids. In particular, a general equation of state is derived for finite closed systems of interacting particles, where the pressure is expressed directly in terms of the density (or system volume) and temperature-volume dependent reducible cluster integrals. Although its accuracy is now greatly affected by the limited character of the existing data on the reducible cluster integrals and, especially, the absence of any information on their density dependence, a number of simple approximations indicate the qualitative adequacy of this equation in various regimes of a fluid: from gaseous to liquid states (including the transition region).
\end{abstract}


\end{frontmatter}


\section{Introduction}

Recent studies of Mayer's cluster expansion (CE) \cite{Mayer,Pathria,Balescu,JML} have significantly advanced the statistical theory of the first-order phase transitions. New equations of state in terms of irreducible cluster integrals (virial coefficients) \cite{PRL,Bannur1,PRE1,JCP1} have clarified the actual limitations of the well-known virial expansion for pressure in powers of density (virial equation of state, VEoS, \cite{Mayer,Pathria,Balescu}) and indicated the beginning of the condensation process at the density, $\rho _G$, where the VEoS isothermal bulk modulus vanishes \cite{PRL,Bannur1}. Studies of the virial expansions for pressure and density in terms of reducible cluster integrals \cite{UPJ4,PRE4,Pramana} (virial series in powers of activity, AVEoS, \cite{Mayer,Pathria}) have established the fact of their divergence at the activity, $z_G$, which corresponds exactly to the same density, $\rho _G$ \cite{Mayer,Bannur1,JML,Pramana}. Moreover, the observed character of this divergence directly indicates the beginning of the condensation: beyond the $\rho _G$, AVEoS yields a jump of density at constant pressure and chemical potential (activity $z_G$) \cite{UPJ4,PRE4,Pramana}.

Thus, the modern statistical theory provides a way to locate the fluid saturation point, $\rho _G$, analytically on the basis of information about the interaction potential (in principle, the reducible and irreducible integrals can be defined on the basis of this information \cite{Mayer,Pathria,JML}), however the location of the corresponding boiling point, $\rho _L$, remains much more difficult to define. As it has been shown in a number of papers \cite{PRL,JML,PRE4}, the CE with constant (density independent) cluster integrals cannot reproduce the true finite jump of density (from $\rho _G$ to $\rho _L$ or vice versa) through the phase transition. All the equations mentioned above (in terms of the constant reducible as well as irreducible integrals) yield an essential discontinuity of density (the divergence to infinity) instead of the proper jump discontinuity. Although the main reason for such non-physical behavior is known in principle (it has been clearly stated in some researches \cite{UPJ4,PRE4,JML}) the problem still remains absolutely unexplored in statistical theory.

For a wide range of lattice-gas models, this problem can simply be avoided due to the recently discovered "hole-particle" symmetry of the binodal \cite{UPJ4,PRE3,PRE2} (i.e., the symmetry between $\rho _G$ and $\rho _L$), but, for continuous statistical models of matter, the corresponding symmetry is not so obvious, and the relation between $\rho _G$ and $\rho _L$ would have much more complex character.

In this paper, the first important steps are made in a possible way to resolve the problem of the CE inadequacy at very dense regimes of model systems (i.e., regimes that correspond to the condensed states of matter). The key points of the proposed solution are discussed in Section 2. Section 3 presents the derivation of a general equation of state (in terms of volume-dependent reducible integrals), which can potentially describe the behavior of a fluid in all regimes: from gaseous to liquid states continuously. Section 4 is devoted to some attempts to roughly approximate the volume-dependence for high-order reducible integrals. This section also presents the results of the corresponding computations and their discussion. The last section emphasizes the key results of the study and highlights the main directions of possible further developments in the area.

\section{Theoretical backgrounds of the problem} \label{sec2}

\subsection{Limitations on the volume-independence of cluster integrals}

\begin{figure}
\centering{
\includegraphics[scale = 0.7]{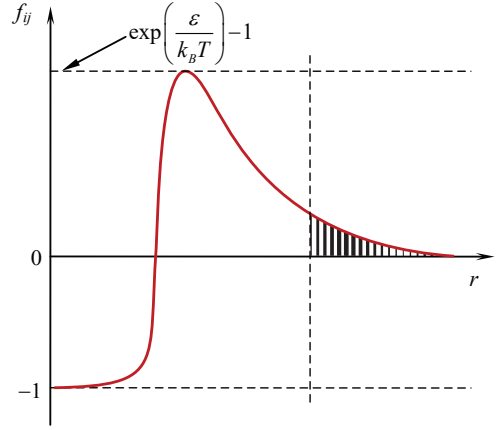}
}
\caption{\label{fig1}Mayer's function for a typical realistic interaction potential ($\varepsilon$ is the depth of the potential well). Any limitations of the integration limits would first affect the attractive (positive) component of the function.}
\end{figure}

The first question that naturally arises, when one has to consider the volume dependence of the cluster integrals, is how these integrals may be affected by changing the integration limits on the macroscopic level (in thermodynamic systems, the number of particles, $N \to \infty $, and the volume, $V \to \infty $ from "a microscopic point of view"), if their integrand (various products of Mayer's functions, see Fig.~\ref{fig1}) vanishes at the microscopic distance between particles.

Indeed, the simplification of the integration limits (their transformation to infinite ones and, hence, the independence of the integrals on the system volume) should be adequate for microscopic (i.e., the low-order, $n \ll N$) clusters even at very dense states while the system volume remains macroscopic. However, this simplification becomes invalid for macroscopic clusters ($n \to N$) at dense states of the system.

This issue is obvious on the example of the well-known expansions for pressure and density in powers of activity (AVEoS):
\begin{equation}
\left. \begin{array}{l}
\frac{P}{{{k_B}T}} = \sum\limits_{n = 1}^\infty  {{b_n}{z^n}} \\
\rho  = \sum\limits_{n = 1}^\infty  {n{b_n}{z^n}} 
\end{array} \right\}, \label{eq1}
\end{equation}
which include the so-called \emph{reducible cluster integrals} $\left\{ b_n \right\} $ \cite{Mayer,Pathria}. Each $b_n$ is the integral over the configuration phase-space of $n$ particles (cluster), which are "connected" to each other by all possible combinations of Mayer's functions (see Fig.~1) but not "connected" with the other particles of the system (its integrand is the sum of all possible products of Mayer's functions, where the index of each particle of the cluster is present at least once).

In fact, the reducible integral for $n$ particles can formally be "reduced" to the sum of various products of \emph{irreducible integrals}, $\left\{ \beta _k \right\}$, which, in turn, correspond to the "biconnected" diagrams (the strict definition for both kinds of integrals as well as the detailed description of complex relationship between them can be found in a number of sources: \cite{Mayer, Balescu, Pathria, JML}):
\begin{equation}
{b_n} = {n^{ - 2}}\sum\limits_{\left\{ {{j_k}} \right\}} {\prod\limits_{k = 1}^{n - 1} {\frac{{{{\left( {n \cdot {\beta _k}} \right)}^{{j_k}}}}}{{{j_k}!}}} } , \label{eq12}
\end{equation}
 where all possible integer sets $\left\{ {{j_k}} \right\}$ must satisfy the condition
\[\sum\limits_{k = 1}^{n - 1} {k{j_k}}  = n - 1.\]

\begin{figure}
\centering{
\includegraphics[scale = 0.7]{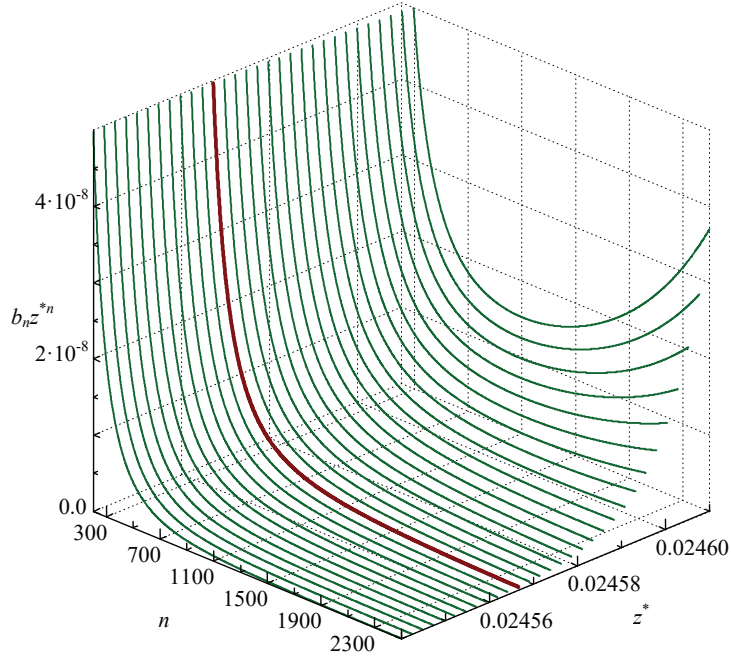}
}
\caption{\label{fig2}Contribution of various terms, $b_{n} {z^*}^{n}$, to the logarithm of the grand partition function [the series for pressure in Eq.~(1)] at various activities (in a dimensionless form, $z^* = z \sigma ^3$) for the Lennard-Jones fluid ($T = 0.9 \varepsilon / k_B$). Cluster integrals are calculated accordingly to \cite{PRE4,JCP3}. The bold line corresponds to the divergence activity, $z_G$.}
\end{figure}

Figure~\ref{fig2} demonstrates the weight (or contribution) of various terms, $b_{n} z^{n}$, to the logarithm of the grand partition function [the series for pressure in Eq.~(\ref{eq1})] and how their weight changes when both the activity and density, $\rho \left( z \right)$, increase (the ranges of the order, $n$, and activity, $z$, are limited in order to visualize the nature of the series divergence at the vicinity of a certain activity, $z_G$). There the reducible integrals, $\left\{ b_n \right\} $, are calculated on the basis of the irreducible ones, $\left\{ \beta _k \right\}$, for the widely known Lennard-Jones model (the high-order $\beta _k$-s are approximated in accordance with \cite{JCP3}) by using the recently derived recursive algorithm \cite{PRE4} which is formally identical to Eq.~(\ref{eq12}).

At dilute states up to the saturation point [while the activity does not exceed the convergence radius, $z \le z_G$, and the particle number density is correspondingly low, $\rho \le \rho _G \left( z_G \right) $], the weight of different terms is a decreasing function of the order, $n$, i.e., the principal contribution to the partition function belongs to the relatively low-order reducible cluster integrals ($n \ll N$). In these regimes, the neglecting of the volume dependence for high-order cluster integrals (and, even, the neglecting of those integrals themselves) does not lead to any loses of accuracy at the thermodynamic limit ($N \to \infty $; $V \to \infty $).
 This fact has recently obtained a strict confirmation \cite{PRE5} for various lattice-gas models and, in particular, the Lee - Yang model \cite{LeeYang}. 

However, the situation turns contra verse at more dense states beyond the saturation point ($\rho > \rho _G$), when the weight of cluster integrals becomes a non-decreasing function in the high-order region (see Fig.~\ref{fig2}). The contribution of the high-order integrals becomes essential there (for infinite series, the contribution tends to infinity for $n \to \infty$ and causes the mathematical divergence) and, at the same time, their volume dependence cannot be neglected in such dense regimes.

The positivity of reducible cluster integrals at subcritical temperatures means that the intermolecular attraction (the positive part of Mayer's function in Fig.~\ref{fig1}) prevails over repulsion (the negative part of Mayer's function in Fig.~\ref{fig1}). Therefore, it is not surprising that the system compression yields the physical phenomenon of condensation at some density high enough (and mathematical divergence of the AVEoS at the vicinity of a certain activity high enough). On the other hand, the volume independence of cluster integrals means that they remain positive at any states (even very dense ones) -- attraction always dominates over repulsion, and, hence, the physical phenomenon of condensation cannot stop (the mathematical divergence is to infinity -- it yields an essential discontinuity of density instead of a jump).

Thus, the observed non-physical behavior of the AVEoS (and other related equations in terms of reducible as well as irreducible cluster integrals) is a direct result of neglecting the actual volume-dependence of cluster integrals (at least, high-order ones).

In real substances, an additional effect can influence the balance between intermolecular attraction and repulsion: the interaction parameters may substantially change in condensed regimes due to the non-additive character of real intermolecular forces. However, this effect is beyond the scope of the present research which is initially focused on statistical models of matter with pair-wise interactions (such as the Lennard-Jones fluid, etc.).

\subsection{Choice of integrals to regard for the volume-dependence} 

Another important issue that has to be considered is what kind of cluster integrals, reducible or irreducible ones, can properly exhibit the real dependence of the partition function on volume in principle.

There is a great experience accumulated for calculating irreducible integrals (i.e., virial coefficients) \cite{Kofke4,UPJ2,Wheatley} for various interaction models \cite{Kofke1,UPJ1,JCP2,UPJ3,Kofke5}, and their volume-dependence would technically be much simpler to explore and apply. Nevertheless, the right choice should be reducible cluster integrals that are rare used by researchers in comparison with irreducible ones, and the other option would be incorrect unfortunately.

As it is mentioned above [see Eq.~(\ref{eq12})], each reducible integral, $b_n$, of a certain order, $n$, is a complex combination (sum of various products) containing the irreducible integrals $\left\{ \beta _k \right\}$ of all possible orders less than $n$ ($k \le {n-1}$). The actual volume-dependence of a high-order reducible integral ($n \to N$) automatically means the volume dependence of all its components: even the irreducible integrals of the lowest orders (for example, $\beta_1$, $\beta_2$, etc.) must depend on the volume as they are parts of the volume-dependent reducible integral of a very high order. On the other hand, each low-order reducible integral ($n \ll N$) is volume-independent in practice, and all its components (i.e. the same $\beta_1$, $\beta_2$, etc.) must not depend on volume too.

To make this issue clear, one may consider the different parts (summands) of some $b_n$ [see Eq.~(\ref{eq12})]: $\beta_{n-1}$, $\beta _1 \beta _{n-2}$, $\beta _1^2 \beta _{n-3}$, $\beta _2 \beta _{n-3}$, ..., $\beta _1^{n-1}$. For a macroscopic cluster ($n \to N$) in dense regimes, the first-order irreducible integrals ($\beta _1$) of \emph{various summands} (even belonging to the same $b_n$) would actually be \emph{different integrals}. As to the irreducible integrals belonging to different clusters (large and small ones), they must differ even more essentially, so that there is no certain $\beta _1$ for the system as whole.

Unfortunately, the irreducible integrals belonging to different reducible ones become \emph{absolutely indistinguishable} when they are used in all the known equations in terms of virial coefficients (the conventional VEoS \cite{Mayer,Pathria,Balescu} or relatively new equations based on the exact generating function \cite{JML,PRL,Bannur1}). 

In these circumstances, the irreducible integrals cannot be used at all, and the corresponding equations mentioned above will always remain inapplicable to condensed states of matter (with rare exceptions like the equations based on the "hole-particle" symmetry for some specific models of matter \cite{PRE2,PRE3}). The only possibility to make the CE behavior adequate in high-density regimes is to use \emph{the equations in terms of reducible cluster integrals as certain functions of volume} (or density), $\left\{ {b_n} {\left( V \right)} \right\}$.

Although the AVEoS formally meets this criterion, the volume-dependence of reducible integrals makes its practical usage almost impossible because the density in Eq.~(\ref{eq1}) is a series with the power coefficients which depend on density in turn. Therefore, the problem solution needs another equation of state in terms of reducible cluster integrals where the activity is excluded as a parameter.
 
\section{Equation in terms of volume-dependent reducible integrals} \label{sec3}

\subsection{Generating function in terms of reducible integrals}

The initial form of Mayer's CE exactly represents the configuration part of the partition function (the so-called configuration integral, $Q_N$) as a complex sum of products of reducible integrals mentioned above, $\left\{ {b_n} \right\}$:
\begin{equation}
{Q_N} = \sum\limits_{\left\{ {{m_n}} \right\}} {\prod\limits_{n = 1}^N {\frac{{{\left({b_n} {V}\right)}^{{m_n}}}}{{{m_n}!}}} } , \label{eq2}
\end{equation}
where all possible integer sets $\left\{ {m_n} \right\}$ must satisfy the condition
\[
\sum\limits_{n = 1}^N {n {m_n}}  = N.
\]

For the canonical ensemble (a closed system with the constant number of particles), the differentiation of the configuration integral directly yields the equation of state:
\begin{equation}
\frac{P}{{{k_B}T}} = {\left[ {\frac{{\partial \left( {\ln {Q_N}} \right)}}{{\partial V}}} \right]_T} = {\frac {1}{Q_N}} \left( {\frac {\partial Q_N}{\partial V }} \right)_T , \label{eq3}
\end{equation}
and the main problem is to obtain an analytical expression for that configuration integral. 

For a known set of reducible integrals, $\left\{ {b_n} \right\}$, Eq.~(\ref{eq2}) formally provides such expression, but its direct usage is too complex in practice (especially for large systems where the $N$ reaches hundreds or thousands). A more convenient way is to use the generating function for the $Q_N$ in Eq.~(\ref{eq2}),
\begin{equation}
F\left( x \right) = \exp \left( {V\sum\limits_{n = 1}^\infty  {{b_n}{x^n}} } \right) = \sum\limits_{i = 0}^\infty  {{Q_i}{x^i}} 
\label{eq4}. 
\end{equation}

Although this function is not new (it was introduced by the Mayers themselves \cite{Mayer}), and its form is much simpler than that of the generating function in terms of irreducible integrals \cite{JML,PRL,Bannur1}, it has not been used explicitly to derive the equation of state in terms of reducible integrals.

In particular, Eq.~(\ref{eq4}) allows defining the $Q_N$ of an arbitrarily high order, $N$, in the corresponding recursive differentiation: 
\begin{equation}
{Q_i} = {\left. {\frac{{{F^{(i)}}\left( x \right)}}{{i!}}} \right|_{x = 0}} = \frac{V}{i}\sum\limits_{n = 1}^i {n{b_n}{Q_{i - n}}} \label{eq5} , 
\end{equation}
where $Q_0 = 1$. 

Mathematically, both the definitions of the $Q_N$ in Eqs.~(\ref{eq2}) and (\ref{eq5}) are absolutely identical, however the last equation is much more convenient for computations.

\subsection{Recursive equation of state}

In order to find the derivative of the $Q_N$ with respect to volume in Eq.~(\ref{eq3}), one can first define the  corresponding derivative of the generating function in Eq.~(\ref{eq4}):
\[
\frac{{\partial F}}{{\partial V}} = \sum\limits_{i = 0}^\infty  {\frac{{\partial {Q_i}}}{{\partial V}}{x^i}}  = F\sum\limits_{n = 1}^\infty  {\left[ {{b_n} + V\left( {\frac{{\partial {b_n}}}{{\partial V}}} \right)} \right]{x^n}}  = \left( {\sum\limits_{n = 1}^\infty  {\left[ {{b_n} + V\left( {\frac{{\partial {b_n}}}{{\partial V}}} \right)} \right]{x^n}} } \right)\left( {\sum\limits_{i = 0}^\infty  {{Q_i}{x^i}} } \right) .
\]

In this expansion, the power coefficient at $x^N$ is
\[
\frac{{\partial {Q_N}}}{{\partial V}} = \sum\limits_{n = 1}^N {\left\{ {{b_n} + V\left[ {\frac{{\partial {b_n}}}{{\partial V}}} \right]} \right\}{Q_{N - n}}}  = \sum\limits_{n = 1}^N {b_n^*\left( V \right){Q_{N - n}}},
\]
where the following designation is used:
\begin{equation}
b_n^*\left( V \right) = {b_n} + V\left[ {\frac{{\partial {b_n}}}{{\partial V}}} \right] \label{eq6} . 
\end{equation}

As a result, Eq.~(\ref{eq3}) is transformed to \emph{the searched equation of state in terms of volume-dependent reducible integrals},

\begin{equation}
\frac{P}{{{k_B}T}} = \frac{{\sum\limits_{n = 1}^N {b_n^*\left( V \right){Q_{N - n}}} }}{{{Q_N}}} \label{eq7} . 
\end{equation}

In comparison with the VEoS and other equations in terms of irreducible integrals (or virial coefficients), the practical usage of Eq.~(\ref{eq7}) is additionally complicated by a laborious stage of determining the set $\left\{ b_n \right\}$ on the basis of a certain already known irreducible integrals, $\left\{ \beta _k \right\}$ [besides the recursive determining the set of $\left\{ Q _i \right\}$ in Eq.~(\ref{eq5})] even in cases when the reducible integrals are considered as constant. On the other hand, Eq.~(\ref{eq7}) meets all the criteria stated in Section~\ref{sec2}: it contains the reducible integrals, which can potentially be volume-dependent, and expresses the pressure as a function of density (actually, it explicitly involves the number of particles, $N$ and the system volume, $V$).

In 1980s, a similar equation was derived in somewhat different (and more complex) manner \cite{Gibbs1, Gibbs2}: generating function (\ref{eq4}) was not directly used in the derivation, and the volume-dependence of reducible integrals was not studied. Unfortunately, the complexity of the equation in combination with technical limitations of the computational equipment at that period imposed substantial restrictions on the size of the studied systems ($N \le 1024$), and the obtained results even caused some doubts for a long time. Although the recent studies of Mayer's CE with constant cluster integrals \cite{PRL,Bannur1,PRE4} have completely confirmed those results the actual behavior of Eq.~(\ref{eq7}) with regard for the volume-dependence of $\left\{ {b_n} \right\}$ has never been explored yet.

\section{Approximation of the volume-dependence} \label{sec4}

\subsection{Phenomenological ansatz for the volume-dependence}

Provided that the set of volume-dependent reducible integrals, $\left\{ {b_n} {\left( V \right)} \right\}$, is completely and exactly known for a certain model of matter, Eq.~(\ref{eq7}) should theoretically be accurate in all possible states of this model: from dilute to condensed regimes. Indeed, this equation does not have any theoretical restrictions on density or temperature at least for models where Mayer's CE remains valid.

At the moment, there is no realistic statistical model (which includes intermolecular attraction as well as repulsion) with the completely known set of cluster integrals. Some truncated sets of irreducible integrals (virial coefficients) have been calculated for widely used interaction models (such as the Lennard-Jones model \cite{Kofke1,Kofke5}, its modifications \cite{UPJ1,JCP2}, Morse \cite{UPJ3} and square-well \cite{KofkeSW} potentials, etc.). There are also a number of approximations for infinite virial sets \cite{JCP3,UPJ3,Kofke3}. For the reducible cluster integrals, the achievements are even more modest: the techniques of calculating the reducible integrals on the basis of irreducible ones still continue to improve \cite{Gibbs1,UPJ4,PRE4,Pramana}.

As to the volume-dependence of those integrals, it has already been mentioned in the previous sections that the problem remains absolutely unexplored in modern physics. There are a number of possible reasons for such ignoring the problem over the years. The calculations of the cluster integrals still involve considerable technical difficulties, and consideration of the volume-dependence can only additionally complicate the calculations. Up to the last years, Mayer's CE did not reach such dense states of systems, where the volume-dependence becomes really essential: there was no need to consider the volume-dependence for low-order cluster integrals, and only recent studies have raised the issue of calculating the relatively high-order integrals.

On the other hand, a simple qualitative analysis of Mayer's function may help to predict the main features of the searched volume-dependence on the corresponding qualitative level. Of course, such considerations does not have a strict ground and can hardly pretend on accuracy in practical calculations, however, they may indicate some important directions of further developments and invigorate future more strict researches.

Formally, there would be no lose of generality and accuracy to express the searched dependence in the following form,
\begin{equation}
b_n \left( V \right) = b_n^0 g \left( n, T, x \right)
\label{eq8},
\end{equation}
where $b_n^0$ is the $n$-th order reducible integral defined over infinite limits; $g \left( n, T, x \right)$ is a certain function of the order, temperature and special variable, $x$, that is related to the real integration limits.

Some key simplifications and inaccuracies may appear in Eq.~(8) due to possible approximations being used in function $g$ or its variable $x$ on the basis of certain analytical or empirical considerations.

\subsection{Approximation on the qualitative level}

The decreasing of integration limits (under compression of the system) must first affect the positive part of Mayer's function (i.e., the long-distance attractive component of intermolecular forces, see Fig.~\ref{fig1}). Therefore, the reducible integrals, which are initially positive (when defined in infinite limits at subcritical temperatures), should decrease or even change the sign when the integration volume decreases.

As a result, the positive set of reducible integrals in dilute regimes may turn to the complex alternating one in very dense regimes -- the primarily attractive contribution to the partition function (that leads to condensation under compression at subcritical temperatures) may turn to the primarily repulsive one (that is similar to the supercritical virial series which cannot yield the condensation). Mathematically, it should stop the density divergence of the AVEoS at very dense states and means the end of the pressure constancy. Physically, it should stop the condensation and means a certain balance between attraction and repulsion in condensed regimes of matter.

Due to the absence of any reliable and accurate information on such complex behavior of reducible integrals, one can suppose that, in average, the reducible integral vanishes (or its positive value significantly decreases) when its integration volume becomes small enough ($V \to V_0$), i.e. the specific volume per particle, $ \frac{V}{n}$ (it is important to distinguish this cluster quantity from the system specific volume, $\frac {V}{N} = \rho^{-1}$), reaches a certain small quantity, $v_0$ (or the cluster density, $\frac {n}{V}$, reaches a certain large quantity $\rho _0$). Furthermore, as the volume-dependence would be essential for the high-order integrals only, and the behavior of various high-order integrals should not differ fundamentally, the $v_0$ may be the same for those integrals and, therefore, can be used as a key parameter in the $x$ variable,
\begin{equation}
x = \frac{{{V_0}}}{V} = \frac{{n{v_0}}}{V} = \frac{n}{N}\frac{\rho }{{{\rho _0}}} \label{eq9}.
\end{equation}

For microscopic clusters ($n \ll N$), $x$ is always very small ($x \to 0$ even in very dense regimes, when $\rho > \rho _0$). For macroscopic clusters ($n \to N$), this variable vanishes ($ x \to 0$) only at dilute (gaseous) states, but $x \simeq 1$ in dense (condensed) regimes. 

Correspondingly, the expected dependence of function $g$ on variable $x$ [see Eq.~(\ref{eq8})] should satisfy the following criteria:

\renewcommand{\labelitemi}{$\bullet$}
\begin{itemize}
\item $ g \left( x=0 \right) = 1$ -- assures that $b_n \to b_n^0$ at dilute states;
\item $ {\left( \frac{\partial{g}}{\partial{x}} \right)}_{x=0} = 0$ -- smoothes the transition from the $b_n$ constancy to the volume-dependence;
\item $ {\left( \frac{\partial{g}}{\partial{x}} \right)}_{x>0} < 0$ -- corresponds to the decreasing of the function.
\end{itemize}

A specific form of function $g$ would influence the results rather quantitatively rather than qualitatively. Thus, two principally different functions are proposed to approximate the $b_n \left( V \right)$ dependence (\ref{eq8}) in order to study the behavior of Eq.~(\ref{eq7}) on the qualitative level:

\begin{equation}
g(x) = \left[ {1 - {x^i}{{\left( {2 - x} \right)}^i}} \right]
\label{eq10};
\end{equation}

\begin{equation}
g(x) = \frac{1}{{1 + {x^i}\exp \left( {x - 1} \right)}}
\label{eq11}.
\end{equation}

\begin{figure}
\centering{
\includegraphics[scale = 0.75]{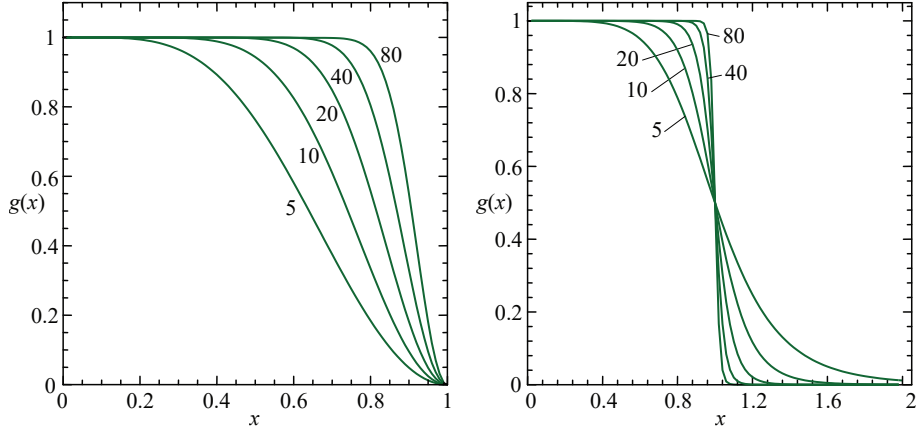}
}
\caption{\label{fig3}Approximating functions (\ref{eq10}) (left) and (\ref{eq11}) (right) for different power factors, $i$ (the numbers in the Figure).}
\end{figure}

The behavior of both functions for various values of the power factor, $i$, is illustrated in Fig.~\ref{fig3}. In general, both functions satisfy the listed above criteria, however the points of their vanishing essentially differ.

\subsection{Results of computations and discussion}

A great number of computations have been performed on the basis of Eq.~(\ref{eq7}) for various interaction models [for the Morse, Lennard-Jones, modified Lennard-Jones models, the truncated sets of known irreducible integrals \cite{Kofke1,UPJ1,JCP2,UPJ3,Kofke5} as well as the sets approximated to infinite orders \cite{JCP3,UPJ3,Kofke3} have been used to calculate the corresponding constant values of reducible integrals $\left\{ b_n^0 \right\}$, see Eq.~(\ref{eq8})] by using both approximations of the $g$ function [see Eqs.~(\ref{eq10}), (\ref{eq11})] in a wide range of their parameters [$i$ and $v_0$, see Eq.~(\ref{eq9})].

\begin{figure}
\centering{
\includegraphics[scale = 0.7]{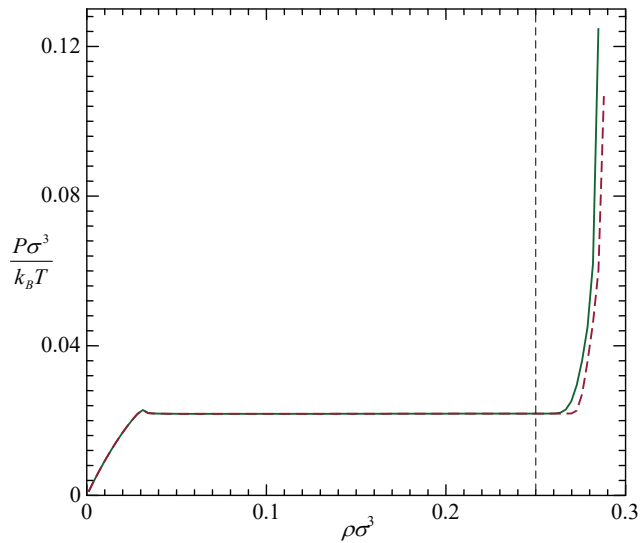}
}
\caption{\label{fig4}Isotherms of Eq.~(\ref{eq7}) for the finite ($N=2000$) Lennard-Jones system ($T = 0.8 \varepsilon / k_B$), where the reducible cluster integrals are volume-dependent according to Eq.~(\ref{eq8}) with approximating functions (\ref{eq10}) (solid line) and (\ref{eq11}) (dotted line). Constant cluster integrals are calculated accordingly to \cite{PRE4, JCP3}. The approximation parameter in Eq.~(\ref{eq9}), $\rho _0 = 0.25 \sigma ^{-3}$ (vertical dotted line).}
\end{figure}

The results of such computations are qualitatively similar: on all the isotherms of Eq.~(\ref{eq7}), the pressure constancy turns into its increasing at some density (which should have the meaning of the boiling-point density) higher than the $\rho _0$ (see Fig.~\ref{fig4}) that well agrees with the all suppositions made above. In fact, the varying of the interaction models and the parameters of approximations affects this behavior quantitatively: only the boiling-point location and the slope of the isotherms beyond this density actually differ in various computations.

It is important to note that the behavior of functions (\ref{eq10}) and (\ref{eq11}) may seem very similar though it fundamentally differs in an important aspect (see Fig.~\ref{fig3}) when their parameter, $i$, is somehow related to the cluster order, $n$: function (\ref{eq10}) vanishes exactly at $x = 1$ for all orders (i.e. all the cluster integrals vanish when their specific volume reaches the same value, $v_0$), but function (\ref{eq11}) vanishes at different $x$ for different orders (i.e. different cluster integrals vanish at different specific volumes). Potentially, such discrepancies could affect the behavior of isotherms in essence. However, the actual difference of that behavior remains qualitative rather than quantitative.

On the other hand, it should additionally be emphasized that the $v_0$ parameter itself [and, hence, its usage in the $x$ variable, see Eq.~(\ref{eq9})] is just a rough simplification: the $v_0$ must be temperature-dependent, and this dependence can principally differ for various clusters. Moreover, the actual dependence of the $x$ variable on the system volume may fundamentally differ from that in Eq.~(\ref{eq9}).

\begin{figure}
\centering{
\includegraphics[scale = 0.7]{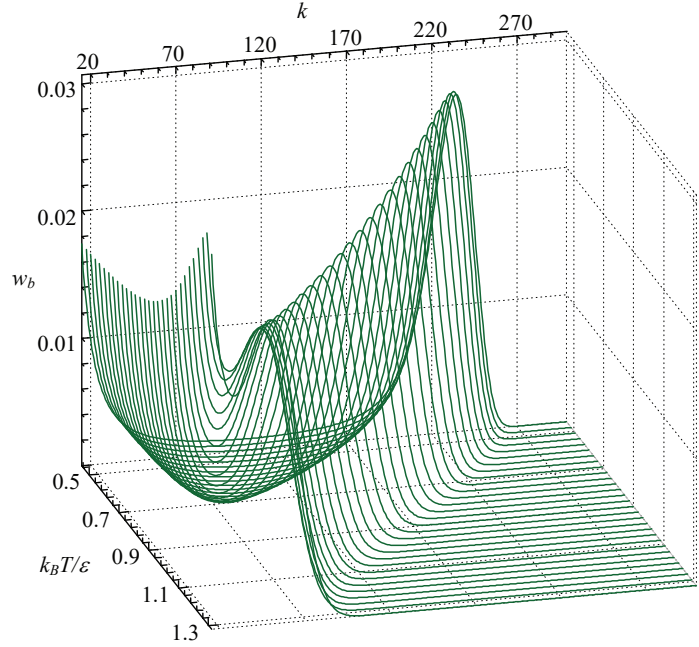}
}
\caption{\label{fig5}Relative weight, $w_b (k)= \left| b_n - b_n^{-}(k) \right| / b_n$, of the irreducible integral, $\beta _k$, in the reducible one, $b_{300}$, [$b_n^{-}(k)$ is the $b_n$ calculated under assumption that $\beta _k = 0$], for various orders and temperatures. Cluster integrals are calculated accordingly to \cite{PRE4, JCP3} for the Lennard-Jones fluid.}
\end{figure}

For example, the real form of the binodal (the liquid branch of all the known binodals, $\rho _L \left( T \right)$, is an essentially non-linear decreasing function of temperature) can be reproduced only if the $v_0$ is a correspondingly increasing function of temperature. Again, this dependence can even be explained qualitatively. As it is mentioned before, each reducible integral is a very complex combination of various "connected" irreducible ones (see detailed description of reducible diagrams in Mayer's book \cite{Mayer}). They may be connected as some "stars", "chains" or their combinations. Figure~\ref{fig5} demonstrates the weights of various irreducible integrals in a certain reducible one at different temperatures (each weight, $w_b \left( k \right)$, is evaluated by comparison of the $b_n$ values calculated with the actual $\beta _k$ and vanishing $\beta _k$). Diagrams of different types cannot be distinguished explicitly there, but some important conclusions may be drawn from this Figure. The high-order irreducible integrals cannot form relatively long "chains": at low temperatures, the contribution of the high-order irreducible integrals is dominant, and the resulting reducible integral rather depends on volume as it is supposed in Eq.~(\ref{eq9}). On the contrary, the low-order irreducible integrals can form "chains" large enough that are much more "sensitive" to any changes of the integration volume: at higher temperatures, the prevailing contribution of the low-order irreducible components makes the reducible integral more sensitive to the system compression that means the corresponding increasing of $v_0$ or even makes Eq.~(\ref{eq9}) inadequate in principle.

\section{Conclusions}

Despite the recent considerable achievements of classical statistics in description of the gas-liquid phase transition on the basis of the cluster-based (Mayer's) approach, the applicability of this approach has still remained extremely limited at condensed states of fluids. The origins of these restrictions are in the volume-independence of the corresponding cluster integrals due to the well-known simplification of their integration limits, when the actual configuration phase-space of a studied system is substituted by the infinite space at the thermodynamic limit.

The present study indicates that the constancy (volume-independence) of the cluster integrals (especially, the high-order ones) does not allow the proper accounting for the intermolecular repulsion in dense supercritical regimes, and this simplification, in turn, causes the non-physical behavior of the existing equations in terms of reducible as well as irreducible cluster integrals (virial coefficients) after the beginning of the condensation process.

A possible way to resolve this problem includes a number of important points. The analysis of Mayer's cluster expansion suggests that, from the two known kinds of cluster integrals, only the reducible ones, $\left\{ {b_n} \right\}$, should be considered as initially volume-dependent, and, on the contrary, the irreducible integrals, $\left\{ {\beta _k} \right\}$, cannot, in principle, be valid functions of the volume (or density) when used in the existing equations. On the other hand, the well-known equation of state with reducible integrals in the form of expansions for pressure and density in powers of activity [AVEoS (\ref{eq1})] cannot technically be used when these integrals are volume-dependent.

Thus, the recursive equation of state in terms of density and reducible cluster integrals has been derived for finite closed systems where these integrals explicitly depend on the system volume [see Eqs.~(\ref{eq6}), (\ref{eq7})]. Although this equation should theoretically be correct in all possible regimes of fluids (in contrast to the previously used equations) its practical usage unfortunately remains hardly possible due to the absence of the reliable and complete information on the high-order cluster integrals and, especially, their actual volume-dependence even for the simplest realistic interaction models.

In order to study the adequacy of Eq.~(\ref{eq7}) at least qualitatively, a number of rough approximations are proposed for the volume-dependence of reducible integrals on the basis of a supposition that the positive value of the subcritical high-order integrals should reduce under the system compression and vanish at some small specific volume per particle for the corresponding cluster.

The computations based on these approximations have reproduced the isotherms that are qualitatively similar to the subcritical isotherms of real substances: the jump of density at constant pressure ends at some dense state, which can be interpreted as a boiling point, and further compression leads to the increasing of pressure that may correspond to the condensed (liquid) regimes of fluids.

It is important to note that the computations mentioned above do not pretend to accuracy as to the actual behavior of the studied model systems (such as the Morse or Lennard-Jones fluids) at condensed states. At the current stage of our knowledge about the cluster integrals, the proposed approximations are too rough to be used in quantitative theoretical description of the condensation phenomenon. In fact, these approximations (and all their parameters as well) do not even have a strict theoretical ground -- they are based only on the logical suppositions about the real behavior of the high-order cluster integrals.

The performed computations rather demonstrate the possibilities of Eq.~(\ref{eq7}) on the qualitative level -- its ability to describe the behavior of fluids continuously from dilute to condensed regimes including the phase transition region.

In essence, the actual reasons for the non-physical behavior of Mayer's cluster expansion at dense states can already be considered as completely known, and the present study clearly establishes the directions to resolve this problem. There are a lot of laborious researches ahead, and the main efforts should be focused on the evaluation of reducible cluster integrals and, especially, their dependence on the real integration limits.

\section*{Acknowledgments}

Funding for the research was provided by a grant from Ministry of Education and Science of Ukraine, No.0117U000348.



\begin{thebibliography}{10}
\expandafter\ifx\csname url\endcsname\relax
  \def\url#1{\texttt{#1}}\fi
\expandafter\ifx\csname urlprefix\endcsname\relax\def\urlprefix{URL }\fi
\expandafter\ifx\csname href\endcsname\relax
  \def\href#1#2{#2} \def\path#1{#1}\fi

\bibitem{Mayer}
J.~E. Mayer, M.~G. Mayer, Statistical Mechanics, 2nd Edition, John Wiley, New
  York, 1977.

\bibitem{Pathria}
R.~K. Pathria, Statistical Mechanics, Butterworth-Heinemann, Oxford, 1997.

\bibitem{Balescu}
R.~Balescu, Equilibrium and Nonequilibrium Statistical mechanics, John Wiley,
  New York, 1975.

\bibitem{JML}
M.~V. Ushcats, L.~A. Bulavin, V.~M. Sysoev, V.~Y. Bardik, A.~N. Alekseev,
  \href{http://www.sciencedirect.com/science/article/pii/S0167732216326381}{Statistical
  theory of condensation — advances and challenges}, Journal of Molecular
  Liquids 224, Part A (2016) 694 -- 712.
\newblock \href
  {http://dx.doi.org/http://dx.doi.org/10.1016/j.molliq.2016.09.100}
  {\path{doi:http://dx.doi.org/10.1016/j.molliq.2016.09.100}}.
\newline\urlprefix\url{http://www.sciencedirect.com/science/article/pii/S0167732216326381}

\bibitem{PRL}
M.~V. Ushcats, Equation of state beyond the radius of convergence of the virial
  expansion, Phys. Rev. Lett. 109 (2012) 040601.
\newblock \href {http://dx.doi.org/10.1103/PhysRevLett.109.040601}
  {\path{doi:10.1103/PhysRevLett.109.040601}}.

\bibitem{Bannur1}
V.~M. Bannur, Virial expansion and condensation with a new generating function,
  Physica A: Statistical Mechanics and its Applications 419~(0) (2015) 675 --
  680.
\newblock \href
  {http://dx.doi.org/http://dx.doi.org/10.1016/j.physa.2014.10.053}
  {\path{doi:http://dx.doi.org/10.1016/j.physa.2014.10.053}}.

\bibitem{PRE1}
M.~V. Ushcats, Adequacy of the virial equation of state and cluster expansion,
  Phys. Rev. E 87 (2013) 042111.
\newblock \href {http://dx.doi.org/10.1103/PhysRevE.87.042111}
  {\path{doi:10.1103/PhysRevE.87.042111}}.

\bibitem{JCP1}
M.~V. Ushcats, Condensation of the lennard-jones fluid on the basis of the
  gibbs single-phase approach, The Journal of Chemical Physics 138~(9) (2013)
  094309.
\newblock \href {http://dx.doi.org/10.1063/1.4793407}
  {\path{doi:10.1063/1.4793407}}.

\bibitem{UPJ4}
M.~V. Ushcats, L.~A. Bulavin, V.~M. Sysoev, S.~Y. Ushcats, Lattice gas
  condensation and its relation to the divergence of virial expansions in the
  powers of activity, Ukrainian Journal of Physics 62~(6) (2017) 533--538.

\bibitem{PRE4}
M.~V. Ushcats, L.~A. Bulavin, V.~M. Sysoev, S.~Y. Ushcats,
  \href{https://link.aps.org/doi/10.1103/PhysRevE.96.062115}{Divergence of
  activity expansions: Is it actually a problem?}, Phys. Rev. E 96 (2017)
  062115.
\newblock \href {http://dx.doi.org/10.1103/PhysRevE.96.062115}
  {\path{doi:10.1103/PhysRevE.96.062115}}.
\newline\urlprefix\url{https://link.aps.org/doi/10.1103/PhysRevE.96.062115}

\bibitem{Pramana}
S.~Y. Ushcats, M.~V. Ushcats, L.~A. Bulavin, O.~S. Svechnikova, I.~L.
  Mykheliev, Asymptotics of activity series at the divergence point, Pramana --
  J. Phys. (2018) (in press).

\bibitem{PRE3}
M.~V. Ushcats, L.~A. Bulavin, V.~M. Sysoev, S.~Y. Ushcats,
  \href{http://link.aps.org/doi/10.1103/PhysRevE.94.012143}{Virial and
  high-density expansions for the lee-yang lattice gas}, Phys. Rev. E 94 (2016)
  012143.
\newblock \href {http://dx.doi.org/10.1103/PhysRevE.94.012143}
  {\path{doi:10.1103/PhysRevE.94.012143}}.
\newline\urlprefix\url{http://link.aps.org/doi/10.1103/PhysRevE.94.012143}

\bibitem{PRE2}
M.~V. Ushcats,
  \href{http://link.aps.org/doi/10.1103/PhysRevE.91.052144}{High-density
  equation of state for a lattice gas}, Phys. Rev. E 91 (2015) 052144.
\newblock \href {http://dx.doi.org/10.1103/PhysRevE.91.052144}
  {\path{doi:10.1103/PhysRevE.91.052144}}.
\newline\urlprefix\url{http://link.aps.org/doi/10.1103/PhysRevE.91.052144}

\bibitem{JCP3}
M.~V. Ushcats, Communication: Low-temperature approximation of the virial
  series for the lennard-jones and modified lennard-jones models, The Journal
  of Chemical Physics 141~(10) (2014) 101103.
\newblock \href {http://dx.doi.org/http://dx.doi.org/10.1063/1.4895126}
  {\path{doi:http://dx.doi.org/10.1063/1.4895126}}.

\bibitem{PRE5}
M.~V. Ushcats, L.~A. Bulavin, Evidence for the first-order phase transition at
  the divergence region of activity expansions, Phys. Rev. E.  (submitted).

\bibitem{LeeYang}
T.~D. Lee, C.~N. Yang,
  \href{http://link.aps.org/doi/10.1103/PhysRev.87.410}{Statistical theory of
  equations of state and phase transitions. ii. lattice gas and ising model},
  Phys. Rev. 87 (1952) 410--419.
\newblock \href {http://dx.doi.org/10.1103/PhysRev.87.410}
  {\path{doi:10.1103/PhysRev.87.410}}.
\newline\urlprefix\url{http://link.aps.org/doi/10.1103/PhysRev.87.410}

\bibitem{Kofke4}
J.~K. Singh, D.~A. Kofke,
  \href{http://link.aps.org/doi/10.1103/PhysRevLett.92.220601}{Mayer sampling:
  Calculation of cluster integrals using free-energy perturbation methods},
  Phys. Rev. Lett. 92 (2004) 220601.
\newblock \href {http://dx.doi.org/10.1103/PhysRevLett.92.220601}
  {\path{doi:10.1103/PhysRevLett.92.220601}}.
\newline\urlprefix\url{http://link.aps.org/doi/10.1103/PhysRevLett.92.220601}

\bibitem{UPJ2}
M.~V. Ushcats, Modification of the mayer sampling method for the calculation of
  high-order virial coefficients, Ukrainian Journal of Physics 59~(7) (2014)
  737--742.
\newblock \href {http://dx.doi.org/10.15407/ujpe59.07.0737}
  {\path{doi:10.15407/ujpe59.07.0737}}.

\bibitem{Wheatley}
R.~J. Wheatley,
  \href{http://link.aps.org/doi/10.1103/PhysRevLett.110.200601}{Calculation of
  high-order virial coefficients with applications to hard and soft spheres},
  Phys. Rev. Lett. 110 (2013) 200601.
\newblock \href {http://dx.doi.org/10.1103/PhysRevLett.110.200601}
  {\path{doi:10.1103/PhysRevLett.110.200601}}.
\newline\urlprefix\url{http://link.aps.org/doi/10.1103/PhysRevLett.110.200601}

\bibitem{Kofke1}
A.~J. Schultz, D.~A. Kofke, Sixth, seventh and eighth virial coefficients of
  the lennard-jones model, Molecular Physics 107~(21) (2009) 2309--2318.
\newblock \href {http://dx.doi.org/10.1080/00268970903267053}
  {\path{doi:10.1080/00268970903267053}}.

\bibitem{UPJ1}
M.~V. Ushcats, Virial coefficients of modified lennard-jones potential,
  Ukrainian Journal of Physics 59~(2) (2014) 172--178.
\newblock \href {http://dx.doi.org/10.15407/ujpe59.02.0172}
  {\path{doi:10.15407/ujpe59.02.0172}}.

\bibitem{JCP2}
M.~V. Ushcats, Modified lennard-jones model: Virial coefficients to the 7th
  order, The Journal of Chemical Physics 140~(23) (2014) 234309.
\newblock \href {http://dx.doi.org/http://dx.doi.org/10.1063/1.4882896}
  {\path{doi:http://dx.doi.org/10.1063/1.4882896}}.

\bibitem{UPJ3}
M.~V. Ushcats, S.~Y. Ushcats, A.~A. Mochalov, Virial coefficients of morse
  potential, Ukrainian Journal of Physics 61~(2) (2016) 160--167.
\newblock \href {http://dx.doi.org/10.15407/ujpe61.02.0160}
  {\path{doi:10.15407/ujpe61.02.0160}}.

\bibitem{Kofke5}
C.~Feng, A.~J. Schultz, V.~Chaudhary, D.~A. Kofke,
  \href{http://scitation.aip.org/content/aip/journal/jcp/143/4/10.1063/1.4927339}{Eighth
  to sixteenth virial coefficients of the lennard-jones model}, The Journal of
  Chemical Physics 143~(4) (2015) 044504.
\newblock \href {http://dx.doi.org/http://dx.doi.org/10.1063/1.4927339}
  {\path{doi:http://dx.doi.org/10.1063/1.4927339}}.
\newline\urlprefix\url{http://scitation.aip.org/content/aip/journal/jcp/143/4/10.1063/1.4927339}

\bibitem{Gibbs1}
E.~Donoghue, J.~H. Gibbs,
  \href{http://link.aip.org/link/?JCP/74/2975/1}{Condensation theory for
  finite, closed systems}, The Journal of Chemical Physics 74~(5) (1981)
  2975--2989.
\newblock \href {http://dx.doi.org/10.1063/1.441420}
  {\path{doi:10.1063/1.441420}}.
\newline\urlprefix\url{http://link.aip.org/link/?JCP/74/2975/1}

\bibitem{Gibbs2}
J.~H. Gibbs, B.~Bagchi, U.~Mohanty,
  \href{http://link.aps.org/doi/10.1103/PhysRevB.24.2893}{Bimodality of
  cluster-size distribution and condensation in a finite lennard-jones system},
  Phys. Rev. B 24 (1981) 2893--2902.
\newblock \href {http://dx.doi.org/10.1103/PhysRevB.24.2893}
  {\path{doi:10.1103/PhysRevB.24.2893}}.
\newline\urlprefix\url{http://link.aps.org/doi/10.1103/PhysRevB.24.2893}

\bibitem{KofkeSW}
H.~Do, C.~Feng, A.~J. Schultz, D.~A. Kofke, R.~J. Wheatley,
  \href{https://link.aps.org/doi/10.1103/PhysRevE.94.013301}{Calculation of
  high-order virial coefficients for the square-well potential}, Phys. Rev. E
  94 (2016) 013301.
\newblock \href {http://dx.doi.org/10.1103/PhysRevE.94.013301}
  {\path{doi:10.1103/PhysRevE.94.013301}}.
\newline\urlprefix\url{https://link.aps.org/doi/10.1103/PhysRevE.94.013301}

\bibitem{Kofke3}
A.~J. Schultz, D.~A. Kofke,
  \href{http://www.sciencedirect.com/science/article/pii/S0378381215301217}{Vapor-phase
  metastability and condensation via the virial equation of state with
  extrapolated coefficients}, Fluid Phase Equilibria 409 (2016) 12 -- 18.
\newblock \href
  {http://dx.doi.org/http://dx.doi.org/10.1016/j.fluid.2015.09.016}
  {\path{doi:http://dx.doi.org/10.1016/j.fluid.2015.09.016}}.
\newline\urlprefix\url{http://www.sciencedirect.com/science/article/pii/S0378381215301217}

\end{thebibliography}

\end{document}